\def\eqref#1{Eq.~(\ref{#1})}
\def\parenref#1{(\ref{#1})}
\newcommand{\barr}{\begin{eqnarray}}
\newcommand{\earr}{\end{eqnarray}}
\newcommand{\beq}{\begin{equation}}
\newcommand{\eeq}{\end{equation}}
\def\gtsim{\hbox{\kern.25em\raise.5ex\hbox{$>$}\kern-.75em\lower.5ex
   \hbox{$\sim$}\kern.25em}}
\def\ltsim{\hbox{\kern.25em\raise.5ex\hbox{$<$}\kern-.75em\lower.5ex
   \hbox{$\sim$}\kern.25em}}
\def\bra#1{\langle #1 |}  
\def\ket#1{| #1 \rangle}  
\documentstyle[preprint,aps,epsfig]{revtex}
\def\chiB{\chi_{_{B}}} \def\Du{\Delta u } \def\Dx{\Delta x } \def\Dt{\Delta t } 
\def\compwav{\lambda_{ \hbox{\tiny C} }}
\def\ordinalth{ ^{\hbox{\small th}} } 
 
\def\Re{\mathop{\rm Re}\nolimits} \def\Im{\mathop{\rm Im}\nolimits}
\def\lslabel#1{\label{#1} \hskip 20pt plus 150pt minus 25pt\penalty-200 \hbox{~~\small [[ #1 ]]}}
\def\lslabel#1{\label{#1}}
\def\lscite#1{\cite{#1}~[~{\small #1}~]}
\def\lscite#1{\cite{#1}}
\begin{document}
\draft
\tightenlines

\title{Limits of fractality: Zeno boxes and relativistic particles}
\author{L. S. Schulman}
\address{Physics Department, Clarkson University, Potsdam, New York 13699-5820, USA}
\date{\today}

\maketitle

\vskip 1 truein

\begin{abstract}
Physical fractals invariably have upper and lower limits for their fractal structure. Berry has shown that a particle sharply confined to a box has a wave function that is fractal both in time and space, with no lower limit. In this article, two idealizations of this picture are softened and a corresponding lower bound for fractality obtained. For a box created by repeated measurements (\`a la the quantum Zeno effect), the lower bound is $\Dx\sim \Dt (\hbar/{mL})$ with $\Dt$ the interval between measurements and $L$ is the size of the box. For a relativistic particle, the lower bound is the Compton wavelength, $\hbar/mc$. The key step in deriving both results is to write the propagator as a sum over classical paths.
\end{abstract}


\narrowtext
\section{Introduction}

Confinement in a box with hard walls makes things rough. Berry \lscite{berry} showed that the wave function of a particle so restricted is nowhere differentiable, neither in time nor in space. Specifically, let a particle be confined to a box, B, in which its dynamics are free but for which its wave function, $\psi$, has Dirichlet boundary conditions on the walls. Let $\psi$ initially be constant in B. Then as a function of time the graph of $\psi$ (e.g., of $\Re\psi$) at a fixed point has dimension 7/4. For fixed time, the graph of $\psi$ is a hypersurface of dimension $D=d+1/2$ where $d$ is the spatial dimension of B. The boundary area of B is assumed finite. (Other cases as well as nongeneric exceptions are also covered in \lscite{berry}.)

In a recent article \lscite{regularize}, Facchi, Pascazio, Scardicchio and I found a way to build a box using the quantum Zeno effect \hbox{(QZE)}. You ``measure" at intervals $\Delta t$ whether a particle is in a region, B. As $\Delta t\to0$ we showed that the particle does not escape, which is the standard \hbox{QZE}. Our main point was that within the box there is free particle dynamics, {\em with Dirichlet boundary conditions on the walls of B}.

Both \lscite{berry} and \lscite{regularize} idealize unnaturally: you don't find perfect boxes in Nature, you don't create wave functions with sharp discontinuities and you don't measure infinitely often. In this article I will show how backing away from one of the idealizations softens the extreme predictions of fractality. Put differently, observed physical fractals generally have upper and lower ranges of validity. What I show here is that when the quantum Zeno box is made penetrable by virtue of finite time intervals between measurements, the fractality softens below a certain distance scale. Specifically, fractality is lost when
\beq
\Dx < \Dt \frac\hbar{mL} \;,
\lslabel{deltax}
\eeq
where $\Delta t$ is the interval between measurements, $m$ is the particle mass and $L$ the box size. This relation is consistent with the results of \lscite{jacobson,kac}, in that a particle confined to a box the size of its Compton wavelength travels in straight lines. 

This leads me to examine a second implicit idealization. Both \lscite{berry} and \lscite{regularize} use the Schr\"odinger equation as a descriptive framework. The path integral of this equation uses Brownian-motion-like paths, rather than those smooth on the Compton wavelength level, as is the case relativistically. Interestingly, it is possible to recover this same Compton wavelength cutoff of fractality starting from our path integral solution of the {\em Schr\"odinger equation}, plus the existence of {\em some} maximal velocity.

The plan of this article is first to recall the results of \lscite{regularize} on how to build a Zeno box, and how to let it leak. Then I will rederive Berry's results in a heuristic way using path integral language. These perspectives will then be combined to show how the leaks in the box truncate the lower scale for fractality. Finally I will show a second kind of fractality breakdown, arising from limitations of speed.

\section{The Zeno box}
A particle of mass $m$ is put in a one-dimensional box of length $L$ and at intervals of time $\Delta t$ it is checked to see whether it is still in the box. The result of \lscite{regularize} is that as $\Delta t \to 0$ the particle remains in the box and within the box evolves under the free Hamiltonian with Dirichlet boundary conditions. This assertion is proved by studying the (combined) propagator for time-$t$ free evolution, followed by projection:
\barr
G(x,t;y) & \equiv & \bra{x} E_B\, U(t) E_B \ket{y}
   =\chiB(x)\bra{x}U(t)\ket{y}\chiB(y)                    \nonumber\\
&=& \chiB(x) \sqrt{\frac{m}{2\pi it
\hbar}}\exp\left[\frac{im(x-y)^2}{2\hbar t}\right]\chiB(y),
\lslabel{Gbox}
\earr
where $E_B$ is the projection operator on the box, $\chiB$ the characteristic function of the set B (so that $E_B$ applied to a state is multiplication by $\chiB(x)$), and $U(t)$ is the operator of unitary time evolution, $\exp(-iHt/\hbar)$, with $H=p^2/2m$.

The use of the Trotter product formula to analyze \eqref{Gbox} is inappropriate. Considered as a potential, $\log\chiB(x)$ is singular.

The method of \lscite{regularize} was to express $G$ in a basis of eigenfunctions of the Dirichlet Hamiltonian (a selection that does {\em not\/} preordain the conclusion) and develop it in an asymptotic short-time expansion. For the box, B, defined by the interval $[0,L]$, the basis (with $Hu_n=E_nu_n$) is
\beq
u_n (x)= \sqrt{\frac2L} \sin\left(\frac{n\pi x}{L}\right) \;, \quad
            E_n=\frac{\hbar^2n^2\pi^2}{2mL^2} \;, \quad n=1,2,\ldots \;.
\eeq
The matrix elements of $G$ are
\barr
G_{mn}(t) =
  \int_0^L dx  \int_0^Ldy \;
   u_m(x) \,
   \sqrt{\frac{m}{2\pi it \hbar}}
        \exp\left[\frac{im(x-y)^2}{2\hbar t} \right] \,  u_n(y)  \;.
\lslabel{Gmn}
\earr
Small-$t$ asymptotic analysis of \parenref{Gmn} yields
\beq
G_{mn}(t) 
  = \delta_{mn} \left( 1 -\frac{it}{\hbar} E_n \right) + O(t^{3/2})\, .
\lslabel{Gsmallt}
\eeq
For $t\to0$ this recovers the free propagator with Dirichlet boundary conditions. For the $N$-projection QZE formulation, showing that the remainder in \eqref{Gsmallt} is 
$O(t^{3/2})$ is sufficient to show that for $t=1/N$, and $N$ projections, one obtains lossless confinement. Moreover, the fact that this turns out to be diagonal in the Dirichlet basis shows this to be the self-adjoint extension of $H$ chosen by the Zeno projection process. \eqref{Gsmallt} can also be used to study leakage, that is, imperfect confinement for finite measurement frequency.

In \lscite{regularize} this result is extended to many dimensions, to complicated boxes, and to Hamiltonians with potentials. But these matters do not concern us here.

The point that I do take is that to create a hard wall, one does not need to be there all the time, just very frequently.

\section{Fractals in a box}

Before getting fractals, we will see how {\em not\/} to get fractals. Although the initial wave function $\psi(x,0)\equiv \frac1{\sqrt L} \chi_{[0,L]}(x)$ is discontinuous (constant in the box, zero outside), if it is allowed to propagate {\em freely}, that is, on the entire line, it is perfectly well behaved.

Using the free propagator (on the entire line), at time-$t$ this wave function becomes
\beq
\psi(x,t)= \frac1{\sqrt L}\int_0^L dy\, g(x-y,\tau)\;,
\hbox{~where}\;
g(u,s)\equiv \frac{e^{iu^2/2s}}{\sqrt{2i\pi s}} 
\hbox{~and}\; 
\tau\equiv \frac{\hbar t}{m}\;.
\lslabel{lineprop}
\eeq
This is clearly a smooth function of $x$. We also find it convenient to look at the derivative of $\psi$, in particular
\beq
\frac{\partial\psi}{\partial x}
       = \frac1{\sqrt L } \Bigl[g(x,\tau)-g(L-x,\tau)\Bigr]   \;,
\lslabel{psiprime}
\eeq
a result derived by changing $\partial_x$ to $\partial_y$ within the integral. Again, this is smooth.

Returning to fractality, Berry's arguments depend on the following property of slowly converging Fourier series. Consider
\beq
f(u)=\sum_m a_m e^{imu} \;,  \hbox{~~with~} |a_m|^2\sim \frac1{|m|^{\beta}} 
\hbox{~~for~} |m|\to\infty \hbox{~~and~} 1<\beta\leq3\;.
\lslabel{fourierseries}
\eeq
If the phases of $\{a_m\}$ are random or pseudorandom, then it follows that the graph of $\Re f$, $\Im f$, and generically $|f|^2$ are fractals of dimension
\beq
D=\frac12(5-\beta) \;.
\lslabel{fractaldim}
\eeq
I will give a heuristic proof of this result, based on a criterion immediately derivable from box-counting definitions of fractal dimension \lscite{deduce}. If the graph of $f(u)$ has dimension $D$, its mean square increment over a small distance $\Du$ satisfies 
\beq
\langle (\Delta f)^2 \rangle \sim (\Du)^{4-2D}  \;.
\lslabel{incrementdim}
\eeq
The square deviation of $f$ is given by
\beq
|f(u+\Du)-f(u)|^2 = 
  \left|\sum_n\sum_m a_m a_n^* e^{iu(m-n)}
                     2\sin\left(\frac{m\Du}2\right)2\sin\left(\frac{n\Du}2\right)\right|
\lslabel{devsum}
\eeq
The average is performed by invoking the ``pseudorandom" phases to eliminate all $n\neq m$ terms in \eqref{devsum}. This leads to
\beq
\langle (\Delta f)^2 \rangle =  4\sum_n |a_n|^2 \sin^2\left(\frac{n\Du}2\right) 
              \sim K\sum_n \frac1{n^\beta} \sin^2\left(\frac{n\Du}2\right)
\lslabel{gestimate}
\eeq
with $K$ a generic constant. For small $\Du$, and up to small errors and irrelevant factors this can be replaced by an integral and rescaled:
\beq
\langle (\Delta f)^2 \rangle 
  = \int_1^\infty \frac{dx}{x^\beta}\sin^2 \left(\frac{x\Du}2\right)
  =\left(\Du\right)^{\beta-1}\int_{\Du}^\infty \frac{dy}{y^\beta}\sin^2 \left(\frac y2\right)
\lslabel{scaledeltaf}
\eeq
Combining Eqs.\ \parenref{scaledeltaf} and \parenref{incrementdim} one obtains the desired result, \eqref{fractaldim}.

The propagator within the box is given as usual \lscite{pibook} by
\beq
G(x,t;y)=\sum_n \frac2L \sin\left(\frac{n\pi x}L\right)\sin\left(\frac{n\pi y}L\right)
            \exp\left(-it\frac{\hbar n^2 \pi^2}{2mL^2}  \right)
\lslabel{Gboxeig}
\eeq
Recall the definition of the Jacobi theta function \lscite{bellman},
\beq
\theta_3(z,s)\equiv \sum_{n=-\infty}^\infty \exp\left(i\pi s n^2 +2izn\right) \;.
\lslabel{thetadef}
\eeq
Shuffling terms in \eqref{Gboxeig} yields
\beq
G(x,t;y)= \frac1{2L} \theta_3\left(\frac\pi{2L}(x-y),-\frac{\pi\hbar t}{2mL^2}\right)
   - \frac1{2L} \theta_3\left(\frac\pi{2L}(x+y),-\frac{\pi\hbar t}{2mL^2}\right)
\lslabel{Gtheta}
\eeq
Several properties of $\theta_3$ now play a role. First it has period $\pi$ and quasiperiod $\pi s$,
\beq
\theta_3(z+\pi,s)=\theta_3(z,s)\;, 
\hbox{~~and~~} 
\theta_3(z+\pi s,s)= e^{-i\pi s-2iz} \theta_3(z,s)  \;.
\lslabel{quasiper}
\eeq
There is a zero of $\theta_3$ at $z=(1+s)\pi/2$, hence, by \eqref{quasiper}, there is an infinity of them. $\theta_3$ also satisfies an identity of far-reaching significance
\beq
\theta_3(z,s)=(-is)^{-1/2}e^{z^2/i\pi s} \theta_3\left(\frac zs,-\frac1s\right)  \;.
\lslabel{fundamental}
\eeq
If $\Im s>0$, $\theta_3$ is clearly an analytic function of $z$. On the line $\Im s=0$ things get interesting and in fact, from \eqref{Gtheta}, this is the case for the physical propagator. In particular, the countable set of zeros now all lie densely on the real line---this for an integral kernel that preserves norm. I have been fascinated by this peculiarity of the propagator for a long time \lscite{pathspin}, and am not surprised that the wave functions propagated by $G$ can be rough.

Applying the fundamental identity, \eqref{fundamental}, to $G$ is straightforward, and a little algebra yields
\beq
G(x,t;y)=\sum_n\Bigl\{ g(x-y-2nL,\tau) -g(x+y-2nL,\tau) \Bigr\}
\lslabel{pathexpansion}
\eeq
with the same $g$ and $\tau$ that were defined in \eqref{lineprop}. This shifts the propagator from an energy expansion to a {\em path\/} expansion, since each term in \eqref{pathexpansion} can be identified with a classical path. The direct path from $y$ to $x$, not bouncing off any wall, is the positive $n=0$ term in \parenref{pathexpansion}. Consider a path from $y$ to $x$ that bounces off the right wall then the left, then the right again, and then reaches $x$. The distance it covers is: $L-y$ (getting to the right wall) plus $2L$ (right to left to right again) plus $L-x$ (getting from the right wall to $x$). This gives $4L-x-y$, the negative $n=2$ term above. This can be visualized by making copies of the box along the real line, with each bouncing path corresponding to an origin point at a displaced image of $y$ (rather like the universal covering space picture in \lscite{pathspin}). The propagator, \parenref{pathexpansion}, could also have been derived by the method of images. Thus each term in \parenref{pathexpansion} satisfies the Schr\"odinger equation and the particular combination taken is that guaranteed to satisfy Dirichlet boundary conditions.

We next establish Berry's result using the path expansion of the propagator. Again we consider a wave function that is initially $1/\sqrt L$ on $[0,L]$, zero elsewhere.
Now, however, its motion is confined to the box and it evolves under the propagator \parenref{pathexpansion}. In evaluating the spatial derivative of the wave function at time-$t$, the identity that took us from \eqref{lineprop} to \eqref{psiprime} is still valid (although for the ``$x+y$" contribution the sign is opposite), and we get
\beq
\frac{\partial\psi}{\partial x}=\frac2{\sqrt L}
    \sum_n  \left\{  g(x-2nL,\tau)-g(x-(2n+1)L)  \right\}
    = \sum_n (-1)^n  g(x-nL,\tau)
\eeq
At this point the reader may be feeling uncomfortable at the display of the derivative of a function known not to possess one, but we sidestep this problem by adding a small negative imaginary part to any convenient quantity, for example $\tau$. Everything is now analytic and our formal manipulations justified.

Our aim is to calculate $\langle |\Delta \psi|^2\rangle$ and use \eqref{incrementdim}. We write
\barr
\Delta \psi &=&\frac{\partial\psi}{\partial x}\Dx 
  = \frac2{\sqrt L} \sum_n (-1)^n  g(x-nL,\tau)\Dx \nonumber\\
   &=&\frac2{\sqrt L} \sum_n (-1)^n  g(x-nL,\tau)
     \frac {g(x+\Dx-nL,\tau)-g(x-nL,\tau)}{\partial g(x-nL,\tau)/\partial x}
\lslabel{ouraim}
\earr
Recalling the definition of $g$, this becomes
\beq
\Delta \psi =\frac{2\tau}{i\sqrt L} \sum_n (-1)^n \frac{g(x-nL,\tau)}{x-nL}
    \left[ \exp\left( i\frac{(x-nL)\Dx}{\tau} +i\frac{(\Dx)^2}{2\tau}  \right)-1\right]
\lslabel{difference}
\eeq
The next step is to take the absolute value squared of $\Delta \psi$, written as a double sum, say over $n$ and $m$. We then argue that ``random phases" make the $n\neq m$ terms drop out and find
\beq
\langle|\Delta \psi|^2\rangle =\frac{2\tau}{\pi L} \sum_n  \frac1{(x-nL)^2}
    \left[ \sin^2\left( \frac{(x-nL)\Dx}{2\tau}\right)\right]
\lslabel{DeltaPsi}
\eeq
This is essentially the same sum as in \eqref{gestimate}, now with $\beta=2$. It gives a fractal dimension $3/2$, which is the correct value.

Note that in \eqref{DeltaPsi} we could let $\tau$ lose its imaginary part {\em before} considering the limit $\Dx\to0$. This is because the series is convergent without $\tau$'s help.

What about the random phase assumption? For sufficiently small $\Dx$ the contributions to the sum come from large $n$. For the square, there are two sums (over ``$n$" and ``$m$"). The terms $g(x-mL,\tau)g(x-nL,\tau)^*$ will have phases on the scale of $(m^2-n^2)L/\tau$, whose remainder modulo $2\pi$ provides the needed pseudo-randomness.

\section{Limit of fractality in the Zeno box}

With the path-based perspective just developed for box fractality, we can see how leaks in the box affect the roughness of the wave function.

In the Zeno box, the particle is contained through repeated localization measurements, implemented at intervals $\Dt$. The propagator for time evolution to time $t$ (with $N=t/\Dt$ projections) is the $N\ordinalth$ power of the propagator of \eqref{Gbox}. As shown in \lscite{regularize}, in the limit $N\to\infty$ this goes over to the box propagator, given for example in the path expansion form, \eqref{pathexpansion}.

Now consider the contributions to the sum in \eqref{pathexpansion} or \parenref{DeltaPsi}. Those with summation index $n$ bounce off the walls about $n$ times, so that if the final time at which the wave function is evaluated is $t$, they are hitting the walls at intervals $t/n$. But these terms should {\em not\/} appear in a finite-$N$ Zeno-box propagator if $n>N$! For in that case, there would be nothing to stop the particle from passing the often-nonexistent boundary. It follows that the only paths that could appear in an approximation for the Zeno-box propagator must satisfy $n\ltsim t/\Dt$.

But now return to \eqref{scaledeltaf}, a special case of which is encountered in \eqref{DeltaPsi}. The bulk of this integral comes from contributions for which $y=\hbox{O}(1)$. Thus, for example, for $\beta=2$ (our case), the range $1/4\leq y\leq4$ includes more than 75\% of the integral. Now ``$y$" in \eqref{scaledeltaf} is $n\Du$, and $\Du$ in \eqref{DeltaPsi} is $L\Dx/\tau$. It follows that the fractality condition is $nL\Dx/\tau\gtsim1$ or $\tau/ L\Dx\ltsim n$.

Recalling the definition of $\tau$ (\eqref{lineprop}), from the last two paragraphs we have the joint requirements
\beq
\frac {\hbar t}{mL\Dx}\ltsim n   \hbox{~~and~~}n\ltsim\frac t{\Dt} \;.
\lslabel{inequal}
\eeq
It follows that $\psi$ can exhibit fractal structure only for
\beq
\Dx\gtsim \frac\hbar{mL}\Dt \;,
\lslabel{minfrac}
\eeq
which is the same as \eqref{deltax}, but stated as a converse.

There is a second way to arrive at this conclusion. Instead of the fractal dimension formula, we make use of the fundamental identity, \eqref{fundamental}. As above, the sum-over-classical-paths expansion, \parenref{pathexpansion}, leads to the condition $n\ltsim t/\Dt$ for the Zeno box, with $\Dt$ the interval between measurements/projections. But we now implement this restriction in the propagator \parenref{pathexpansion} by means of a cutoff---a gradual one, achieved by making $t$ complex \lscite{tcomplex}. Recall that each term in \eqref{pathexpansion} is of the form $[K/{\sqrt{t}}] \exp \left( {im(\xi -2nL)^2}/{2\hbar t} \right)$, with $K$ a constant and $\xi=x\pm y$. Letting $t\to t-i\epsilon$ with $\epsilon>0$ introduces the real factor
\beq
\exp\left(\frac{-\epsilon m(\xi -2nL)^2}{2\hbar t^2}\right) \sim 
\exp\left(\frac{-2\epsilon mn^2L^2}{\hbar t^2}\right)  \;.
\lslabel{eachterm}
\eeq
To cut off the sum in \parenref{pathexpansion} at some $N$ ($=t/\Dt$), set the argument of the last exponent in \eqref{eachterm} to 1, leading to 
\beq
\epsilon_{_N} \sim \frac{\hbar t^2}{2mN^2L^2}  \;.
\lslabel{epsilonvalue}
\eeq
The next step is to go back to the eigenfunction expansion by means of a second application of \eqref{fundamental}. The result is just the formula \parenref{Gboxeig}, except that $t$ is complex. In particular the argument of the exponent giving the time dependence acquires a real part:
\beq
\left|\exp\left(-it\frac{\hbar n^2 \pi^2}{2mL^2}\right)\right|
           =  \exp\left(-\epsilon_{_N}\frac{\hbar n^2 \pi^2}{2mL^2}\right)
           =  \exp\left(-  \frac{n^2\pi^2\hbar^2 t^2}  {4m^2N^2L^4} \right) \;.
\lslabel{realeigexp}
\eeq
Note the distinction: the lower case $n$ is now the index in the eigenfunction expansion, 
while $N$ is the number of measurement projections. From \eqref{realeigexp} we see that the cutoff in the path summation around $N$ leads to a cutoff in the eigenfunction expansion around
\beq
n=\frac{2mL^2}{\pi\hbar t} N  \;.
\lslabel{cutoffn}
\eeq
Eigenfunctions with greater $n$ will have reduced amplitude. Recalling the form of the eigenfunction, $\sin\left({n\pi x}/{L}\right)$, we see that the cutoff will suppress detailed variation of the wave function on a scale $\Dx\sim L/n\pi$. Substituting the value of $n$ given in \eqref{cutoffn} yields 
\beq
\Dx\sim \frac {\hbar t}{2mLN} =\frac{\hbar}{2mL}\Dt   \;.
\lslabel{otherdx}  
\eeq
Aside from an irrelevant factor of 2, this is the same as our earlier result.

\section{Relativistic limitations}

For relativistic particles it is known that the agitated Brownian-motion-like particle paths characteristic of the path integral smooth out into straight lines. The path expansion, \eqref{pathexpansion}, will allow us to see how this destroys fractality.

\subsection{At the Compton wavelength scale}

A particle of mass $m$ has a characteristic length, its Compton wavelength, $\compwav \equiv \hbar/{mc}$. Rewriting \eqref{minfrac} in terms of this quantity yields
\beq
\frac{\Dx}{\compwav} \gtsim \frac {\Dt}{L/c}
\lslabel{comptonineq}
\eeq
for the minimal level of fractal structure. For the right hand side of this inequality to be small, the interval between measurements would have to be small compared to the time it takes light to cross the box itself. Even without claiming to understand the measurement process, this would seem a limiting time interval, which in turn implies that the particle does not have fractal structure below the scale of its Compton wavelength. 

It would thus appear that the QZE plus a causality-like condition are enough to fix $\compwav$ as a lower level of fractality. As we will see, the result is more general and for box-fractality is more the result of the path expansion than of the QZE.

\subsection{The checkerboard path integral}

These considerations bring to mind properties of relativistic particles that are known from another path integral, the Feynman ``checkerboard" sum for the Dirac equation. In his early work \lscite{schweber}, Feynman developed a sum over paths forumulation for the $1+1$ dimensional Dirac equation. This path integral, independently discovered by Riazanov \lscite{riazanov}, was published by Feynman in \lscite{feynmanhibbs}. In parallel, a sum over paths for the telegrapher equation was proposed by Kac \lscite{mobil}. It was later found \lscite{kac} that this is an analytic continuation of the checkerboard sum and that the underlying wave equations (Dirac and telegrapher) bear the same relation \lscite{history}. These formulations have been developed by many researchers over the years \lscite{jacobson,kac,othercheckerboard}, have been extended in various ways to 3-dimensions \lscite{3dcheckerboard}, and applied to other problems as well \lscite{otherproblems}.

In \lscite{jacobson,kac}, using the checkerboard formulation, we found that at the relativistic level particle trajectories become smooth. Instead of the nonrelativistic path integral's Brownian-motion paths, fractal objects with $\Dx\sim\sqrt{\Dt}$, the particle travels in straight lines at the velocity of light, but at random times reverse direction. These reversals are Poisson distributed with rate $mc^2/\hbar$. This implies that the particle reverses after traveling on the order of one Compton wavelength \lscite{disclaimer}.

The nonrelativistic limit of this, and the emergence of jerky, fractal paths, now becomes a standard piece of mathematics---modulo the usual ``$i$." For fixed distances, you allow the time scale to grow. Consider the physical Brownian motion of say, pollen. A reasonable model is to assume that between collisions the path is a straight line and that the collisions themselves are Poisson distributed. Brownian motion then emerges as a limit (and then an idealization) of this phenomenon. A quantitative version of this transition can be implemented for the nonrelativistic limit of the checkerboard path integral. Our picture, described above, is ``straight line motion at velocity $c$, with reversals on the average once per $\hbar/mc^2$"---leading to an expected distance covered between reversals of $\hbar/mc$. Now the diffusion constant derived from microscopic models of Brownian motion is on the order of $D\sim(\Dx)^2/\Dt$. Putting this together for the checkerboard process, one gets
\beq
D\sim \frac{(\hbar/mc)^2}{\hbar/mc^2} = \frac\hbar m \;.
\lslabel{diffusion}
\eeq
From many considerations \lscite{considerations}, this is the right diffusion constant to assign to Schr\"odinger particles.

From \lscite{kac}, the condition for the validity of this procedure is $mc^2/\hbar \gg c\partial_x$. This inequality comes from a proposed expansion of the propagator, so that the characteristic momentum would be $m\delta x/\delta t$ (with $\delta x$ and $\delta t$ increments of the arguments of the propagator). Replacing $\partial_x$ by $m\delta x/\hbar\delta t$, the quoted condition becomes $\delta x\ll c\delta t$.

\subsection{Fractality breakdown from the existence of a limiting velocity}

Starting from the Schr\"odinger equation the same conclusions about smoothing can be drawn from the mere existence of a limiting velocity, $c$. Explicitly the expansion of the box propagator, \eqref{pathexpansion}, is
\beq
G(x,t;y)=\sum_n \sqrt{\frac{m}{2\pi it \hbar}}
\left\{   \exp\left[\frac{im(x-y-2nL)^2}{2\hbar t}\right]
       -   \exp\left[\frac{im(x+y-2nL)^2}{2\hbar t}\right]
 \right\}   \;.
\lslabel{boxsum}
\eeq
For large $n$, the $n\ordinalth$ path has covered a distance $2nL$, and has done it in time $t$. The existence of a limiting velocity implies that such terms should not appear if $2nL>ct$. In other words, we only allow paths with $n<{ct}/(2L)$ By the same arguments that followed \eqref{inequal} a cutoff in $n$ in the integral \parenref{scaledeltaf} implies $n\geq \hbar t /(mL\Dx)$. Combining this with the constraint just derived from the existence of a limiting velocity, we have
\beq
\frac{ct}{2L} > n \gtsim \frac{\hbar t}{Lm\Dx} \;,
\eeq
implying
\beq
\Dx \gtsim \frac\hbar{mc}  \;.
\lslabel{speedlimit}
\eeq
A limiting velocity implies a limit on fractal structure. If $c$ is identified as physical light velocity, the limit is the Compton wavelength.

\acknowledgments
I thank Paolo Facchi, Saverio Pascazio and Antonello Scardicchio for lively discussions. This work was supported by the United States National Science Foundation Grants PHY 97 21459 and PHY 00 99471.



\begin{references}

\bibitem{berry} M. V. Berry, Quantum fractals in boxes, {\it J. Phys. A} {\bf 29}, 6617 (1996).

\bibitem{regularize} P. Facchi, S. Pascazio, A. Scardicchio and L. S. Schulman, Zeno dynamics yields ordinary constraints, {\em Phys.\ Rev.\ A}, to appear.

\bibitem{jacobson} T. Jacobson and L. S. Schulman, Quantum Stochastics: The Passage from a Relativistic to a Non-Relativistic Path Integral, {\it J. Phys.\ A} {\bf 17}, 375 (1984).

\bibitem{kac} B. Gaveau, T. Jacobson, M. Kac and L. S. Schulman, Relativistic Extension of the Analogy between Quantum Mechanics and Brownian Motion, {\it Phys.\ Rev.\ Lett.} {\bf 53}, 419 (1984).

\bibitem{deduce} Formula \parenref{incrementdim} is itself easily deducible. Using a measuring box of linear dimension $\Du$, let the length of a curve be $N(\Du)\Du$, with $N(\Du)$ the number of boxes needed to cover the curve. The fractal dimension is then \lscite{mandelbrot} the limit of $-\log N(\Du)/\log\Du$. From \eqref{incrementdim}, the length of the graph of the function is $\sqrt{\langle(\Delta f)^2\rangle}/\Du$.

\bibitem{pibook} L. S. Schulman, {\it Techniques and Applications of Path Integration}, Wiley, New York (1981) (reissued, ``Wiley Classics Library," 1996).

\bibitem{bellman} R. Bellman, {\it A Brief Introduction to Theta Functions}, Holt, Rinehart and Winston, New York (1961).

\bibitem{pathspin} L. S. Schulman, A Path Integral for Spin, {\it Phys.\ Rev.} {\bf 176}, 1558 (1968).

\bibitem{tcomplex} Any other convenient quantity could also have been made complex, for example $m$ or~$\hbar$.

\bibitem{schweber}S. S. Schweber, Feynman and the visualization of space-time processes, {\em Rev.\ Mod.\ Phys.} {\bf 58}, 449 (1986).

\bibitem{riazanov} G. V. Riazanov, The Feynman Path Integral for the Dirac Equation, {\em Sov.\ Phys.\ JETP} {\bf 6}, 1107 (1958).

\bibitem{feynmanhibbs} R. P. Feynman and A. R. Hibbs, {\em Quantum Mechanics and Path Integrals}, McGraw-Hill, New York (1965).

\bibitem{mobil} M. Kac, {\em Some Stochastic Problems in Physics and Mathematics},
Field Research Laboratory Socony Mobil Oil Co., Inc. Colloquium lectures in the Pure and Applied Sciences (1956); M. Kac, A Stochastic Model Related to the Telegrapher's Equation, {\em Rocky Mount.\ J. Math.} {\bf 4}, 497 (1974).

\bibitem{history} There is an interplay of human and scientific connections here. It can be formulated in diagrammatic form:
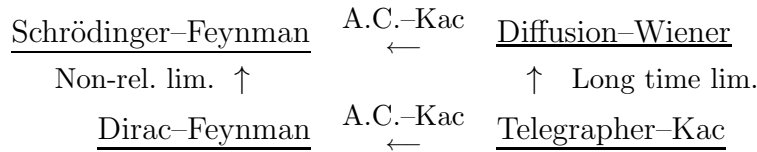
\barr
\hbox{\underline{Schr\"odinger--Feynman}~~}&
   { {\hbox{\small A.C.--Kac}} \atop {\longleftarrow}} &
            \hbox{~~\underline{Diffusion--Wiener}} \nonumber \\
\hbox{\small Non-rel.\ lim.} ~ \uparrow ~~~~~~~ &  ~~~~~~~~~~~~~~&
    ~~~~ \uparrow  ~~ \hbox{\small Long time lim.}   \nonumber \\
\hbox{\underline{Dirac--Feynman}~~}& 
   { {\hbox{\small A.C.--Kac}} \atop {\longleftarrow}} &
        \hbox{~~\underline{Telegrapher--Kac}} \nonumber
\earr
where ``A.C." stands for ``Analytic continuation." Each underlined entry represents a wave equation and the individual identified with developing a path summation formula for it. The top line is obtained from the bottom line by a limiting procedure and the right side goes over to the left (roughly) by analytic continuation, in both cases discovered by Kac.

\bibitem{othercheckerboard} H. A. Gersch, Feynman's Relativistic Chessboard as an Ising Model, {\em Int.\ J. Theor.\ Phys.} {\bf 20}, 491 (1981); G. N. Ord, A Reformulation of the Feynman Chessboard Model, {\em J. Stat.\ Phys.} {\bf 66}, 647 (1992).

\bibitem{3dcheckerboard}B. Gaveau and L. S. Schulman, Grassmann-valued processes for the Weyl and the Dirac equations, {\em Phys.\ Rev.\ D} {\bf 36}, 1135 (1987); B. Gaveau and L. S. Schulman, Dirac Equation Path Integral: Interpreting the Grassmann Variables, {\em Nuovo Cim.\ D} {\bf 11}, 31 (1989); T. Jacobson, Spinor chain path integral for the Dirac equation, {\em J. Phys. A} {\bf 17}, 2433 (1984); G. N. Ord and D. G. C. McKeon, On the Dirac Equation in 3+1 Dimensions, {\em Ann.\ Phys.} {\bf 222}, 244 (1993).

\bibitem{otherproblems} B. Gaveau and L. S. Schulman, Charged Polymer in an Electric Field, {\em Phys. Rev. A} {\bf 42}, 3470 (1990); D. Mugnai, A. Ranfagni and R. Ruggeri, Path-Integral Solution of the Telegrapher Equation:  An Application to the Tunneling Time Determination, {\em Phys.\ Rev.\ Lett.} {\bf 68}, 259 (1992); C. DeWitt-Morette and S. K. Foong, Path integral solutions of wave equations with dissipation, {\em Phys.\ Rev.\ Lett.} {\bf 62}, 2201 (1989); L. H. Kauffman, H. P. Noyes, Discrete physics and the Dirac equation, {\em Phys.\ Lett.\ A} {\bf 218}, 139 (1996).

\bibitem{disclaimer} My characterization of particle motion requires a (philosophical?) disclaimer. In writing ``the particle travels in straight lines," what is meant is that there is a probabilistic model of the wave equation in which the solution of the equation can be arrived at by summing over paths of this sort.

\bibitem{considerations} The most obvious such assignment comes from comparison of the the Schr\"odinger equation ($i\partial_t \psi = (\hbar/2m)\partial^2_x\psi$) and the diffusion equation ($\partial_t \rho = D\partial^2_x \rho$). More direct ties have been proposed by a number of authors, among them: E. Nelson, Derivation of the Schr\"odinger Equation from Newtonian Mechanics, {\em Phys.\ Rev.} {\bf 150}, 1079 (1966); M. Roncadelli, Classical dynamical origin of Feynman paths? {\em J. Phys.\ A} {\bf 26}, L949 (1993); M. Roncadelli, New path integral representation of the quantum mechanical propagator, {\em J. Phys.\ A} {\bf 25}, L997 (1992); M. Roncadelli and A. Defendi, {\em I Camini di Feynman}, Univ.\ degli Studi di Pavia, Pavia, Italy (1992).

\bibitem{mandelbrot} B. B. Mandelbrot, {\em The Fractal Geometry of Nature}, Freeman, New York (1983); J. Feder, {\em Fractals}, Plenum Press, New York (1988).

\end{references}
\end{document}